\documentclass[]{emulateapj}

\begin{document}

\slugcomment{ApJ Submitted}

\newcommand{\kms}{\hbox{km s$^{-1}$}}
\newcommand{\msun}{$M_{\odot}$}
\newcommand{\rsun}{$R_{\odot}$}
\newcommand{\vsini}{$v\sin{i}$}
\newcommand{\teff}{$T_{\rm eff}$}
\newcommand{\logg}{$\log{g}$}
\newcommand{\tarr}{$t_{\rm arr}$}
\newcommand{\mas}{mas yr$^{-1}$}

\title{ MMT Hypervelocity Star Survey III:  A Complete Survey of Faint B-type 
Stars in the Northern Milky Way Halo }

\author{Warren R.\ Brown,
        Margaret J.\ Geller, and
        Scott J.\ Kenyon}

\affil{ Smithsonian Astrophysical Observatory, 60 Garden St, Cambridge, MA 02138 }

\email{wbrown@cfa.harvard.edu}

\shorttitle{ MMT Hypervelocity Star Survey III }
\shortauthors{Brown, Geller, \& Kenyon}

\begin{abstract}

	We describe our completed spectroscopic survey for unbound hypervelocity
stars (HVSs) ejected from the Milky Way.  Three new discoveries bring the total
number of unbound HVSs to 21.  We place new constraints on the nature of HVSs and on
their distances using moderate resolution MMT spectroscopy.  Half of the HVSs are
fast rotators; they are certain 2.5 - 4 \msun\ main sequence stars at 50 - 120 kpc
distances.  Correcting for stellar lifetime, our survey implies that unbound 2.5 - 4
\msun\ stars are ejected from the Milky Way at a rate of $1.5\times10^{-6}$
yr$^{-1}$.  The observed HVSs are likely ejected continuously over the past 200 Myr
and do not share a common flight time.  The anisotropic spatial distribution of HVSs
on the sky remains puzzling.  Southern hemisphere surveys like SkyMapper will soon
allow us to map the all-sky distribution of HVSs.  Future proper motion measurements
with {\it Hubble Space Telescope} and {\it Gaia} will provide strong constraints on
origin.  All existing observations are consistent with HVS ejections from encounters
with the massive black hole in the Galactic center.

\end{abstract}


\section{INTRODUCTION}

	HVSs are unbound stars escaping the Milky Way.  \citet{hills88} first
predicted their existence as a consequence of the massive black hole (MBH) in the
Galactic center.  Present-day observations provide compelling evidence for a
$4\times10^6$ \msun\ central MBH, surrounded by an immense crowd of stars
\citep{gillessen09, do13}.  Theorists estimate that three-body interactions with
this MBH will unbind stars from the Galaxy at a rate of $\sim$10$^{-4}$ yr$^{-1}$
\citep{hills88, yu03, perets07, zhang13}.  The ejection rate of unbound stars by the
central MBH is 100$\times$ larger than the rate expected by competing mechanisms,
including unbound runaway ejections from the Galactic disk \citep{brown09a,
perets12}.

	\citet{brown05} serendipitously discovered the first HVS in the outer
stellar halo, a B-type star moving over twice the Galactic escape velocity.  This
discovery motivated our targeted HVS Survey with the MMT telescope.  The HVS Survey
has identified at least 16 unbound stars over the past seven years \citep{brown06,
brown06b, brown07a, brown07b, brown09a, brown09b, brown12b}.  Other observers have
found unbound and candidate unbound stars among a range of stellar types
\citep{hirsch05, edelmann05, heber08, tillich09, irrgang10, tillich11, li12}. The
variety of HVS observations has led to some confusion.

	HVSs are rare objects.  Of the Milky Way's $10^{11}$ stars, there should be
only 1 HVS within 1 kpc of the Sun for an HVS ejection rate of $10^{-4}$ yr$^{-1}$.  
Thus the vast majority of fast-moving stars near the disk are disk runaways
\citep{bromley09}, not HVSs ejected by the MBH. \citet{heber08}'s discovery of the
first unbound B star ejected from the disk is a case in point.  The Milky Way's
stellar halo contains a millionfold more normal stars than HVSs, as demonstrated by
the absence of unbound A- and F-type stars in the Sloan Digital Sky Survey
\citep{kollmeier07, kollmeier09, kollmeier10}.  In this context, metal poor F-type
stars with marginally unbound proper motion velocities are probably halo stars.

	The HVS Survey is a clean, well-defined spectroscopic survey of stars with
the colors of 2.5 - 4 \msun\ stars.  These stars should not exist at faint
magnitudes in the outer halo unless they were ejected there.  The stars we define as
HVSs are significantly unbound in radial velocity alone.  To date, all follow-up
observations show that the HVSs are main sequence B stars at 50 - 100 kpc distances
\citep{fuentes06, przybilla08b, lopezmorales08, brown12c, brown13b}.  B-type stars
have relatively short lifetimes and must originate from a region with on-going star
formation, such as the Galactic disk or Galactic center.  For those HVSs with
detailed echelle spectroscopy, their stellar ages exceed their flight times from the
Galaxy by $\simeq$100 Myr, an observation difficult to reconcile with Galactic disk
runaway scenarios involving massive stars \citep{brown12c, brown13b}.

	Here we describe the completed MMT HVS Survey.  In Section we begin 2 by
describing our data and the discovery of 3 new HVSs.  One of these HVSs is in the
south Galactic cap where there are few other HVS discoveries to date.  In Section 3
we describe our stellar atmosphere fits to the HVS spectra, and identify rapidly
rotating HVSs that are certain main sequence B stars at 50 - 120 kpc distances.  In
Section 4 we investigate the flight time distribution of HVSs and find that the full
sample of HVSs is best described by a continuous distribution.  The eleven HVSs
clumped around the constellation Leo have flight times equally well described by a
burst or a continuous distribution.  In Section 5 we predict proper motions for
allowable Galactic disk and Galactic center ejection origins, and show that future
proper motion measurements with {\it Hubble Space Telescope} and {\it Gaia} can
distinguish between these origins.  Finally, in Section 6 we use the completed HVS
sample to estimate a $1.5\times10^{-6}$ yr$^{-1}$ ejection rate of unbound 2.5 - 4
\msun\ stars from the Milky Way.

\section{DATA}

\subsection{Target Selection}

	The HVS Survey is a spectroscopic survey of late B-type stars selected by
broadband color \citep{brown06}.  The final color selection is detailed in
\citet{brown12b} and we do not repeat it here.  Photometry comes from the Sloan
Digital Sky Survey \citep[SDSS,][]{aihara11}.  We correct all photometry for
reddening following \citet{schlegel98} and exclude any region with high reddening
$E(B-V) > 0.1$ mag.  We also exclude two small end-of-stripe regions with bad
colors, and all objects within 2\arcdeg\ of M31.  The revised list contains 1451 HVS
Survey candidates.

	The HVS Survey color selection primarily identifies evolved blue horizontal
branch (BHB) stars in the stellar halo, but it also identifies some $\simeq$3 \msun\
main sequence B stars.  BHB stars and $\simeq$3 \msun\ main sequence B stars share
similar effective temperatures and surface gravities in the color range of the HVS
Survey, and both BHB and main sequence B stars are intrinsically luminous objects
($g$-band absolute magnitudes $-1<M_g<1$ mag).  Because the HVS Survey covers high
Galactic latitudes $|b|\gtrsim30\arcdeg$, the $17<g_0<20.25$ magnitude-limited HVS
Survey exclusively targets stars in the stellar halo of the Milky Way.

\subsection{Spectroscopic Observations and Sample Completeness}

	We obtain spectra for 269 new HVS Survey candidates.  We also obtained
repeat observations of previously identified HVSs to validate their nature and
radial velocities.  New observations were acquired at the 6.5m MMT Observatory in
five observing runs between April 2012 and October 2013.  All observations were
obtained with the Blue Channel Spectrograph \citep{schmidt89} using the 832 line
mm$^{-1}$ grating and the 1\arcsec\ slit.  This set-up provides a wavelength
coverage of 3650 \AA\ -- 4500 \AA\ at a spectral resolution of 1.0 \AA.  We chose
exposure times to yield a signal-to-noise ratio (S/N) of 10 to 15 per resolution
element in the continuum.  All observations were paired with a He-Ne-Ar lamp
exposure for accurate wavelength calibration, and were flux-calibrated using blue
spectrophotometric standards \citep{massey88}.

	We have now obtained spectra for 1435 of the 1451 HVS Survey candidates.
This count includes 63 objects with spectra taken from SDSS.  The HVS Survey
completeness is thus 99\%.  In our sample we find 255 (18\%) hydrogen atmosphere
white dwarfs, 24 (2\%) quasars, and 30 (2\%) miscellaneous objects such as metal
poor galaxies \citep{kewley07} and B supergiants \citep{brown07c, brown12b}.  
Low-mass white dwarfs are a problematic contaminant:  they can appear as velocity
outliers because of binary orbital motion \citep{kilic07b}.  We have thus
systematically identified and removed the low mass white dwarfs for study elsewhere
\citep{kilic10, kilic11a, kilic12a, brown10c, brown12a, brown13a}.  Our focus here
is the cleanly selected sample of 1126 late B-type stars.  This complete
spectroscopic sample is the basis for the following analysis.

\begin{figure}          
 \plotone{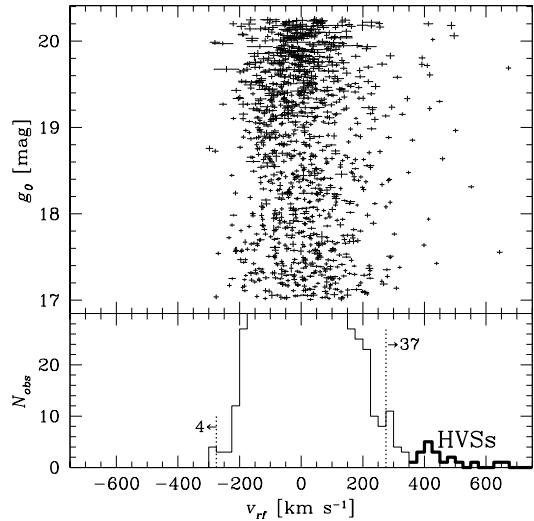}
 \caption{ \label{fig:velh}
        Distribution of Galactic rest-frame velocity and de-reddened $g$-band
magnitude for the 1126 late B-type stars in the HVS Survey.  Stars with velocity
errors greater than 20 \kms\ are mostly SDSS measurements.  The lower panel plots
the velocity histogram emphasizing the tails of the distribution (the histogram
peaks at $N_{obs}=104$ in this binning).  The significant lack of $v_{rf}<-275$
\kms\ stars demonstrates that the majority of bound positive velocity outliers have
lifetimes less than the orbital turn-around time.  Stars with $v_{rf}>400$ \kms\ are
unbound.}
 \end{figure}

\subsection{Radial Velocity Distribution}

	The goal of the HVS Survey is to find velocity outliers.  Having completed
the HVS Survey, we re-measured the stellar radial velocities from all of our spectra
using the latest version of the cross-correlation package RVSAO \citep{kurtz98}.  
As in earlier work, we use high S/N spectra of bright late B- and early A-type
radial velocity standards \citep{fekel99} as our cross-correlation templates.  The
mean radial velocity uncertainty of our measurements is $\pm$10 \kms.

	For the 63 objects observed by SDSS we adopt the radial velocity reported by 
the SEGUE Stellar Parameter Pipeline \citep{allende08, smolinski11}.  The mean 
radial velocity uncertainty of the 63 stars observed by SDSS is $\pm$22 \kms.

	Our primary interest is not heliocentric velocity, but velocity in the 
Galactic rest frame.  To properly interpret the radial velocities of stars in the 
halo requires that we correct for the local circular velocity and the motion of the 
Sun with respect to the disk.  We have long assumed a circular velocity of 220 \kms, 
but in \citet{brown12b} we adopted a circular velocity of 250 \kms\ on the basis of 
disk maser proper motions \citep{reid09, mcmillan10}.  Subsequently, \citet{bovy12} 
measured a circular velocity of 220 \kms\ on the basis of stellar radial velocities 
in the inner halo, but a Local Standard of Rest motion 14 \kms\ larger than 
previously measured.  These different measurements reconcile with each other for a 
circular velocity of 235 \kms\ and the Local Standard of Rest motion of 
\citet{schonrich10}, which we adopt here.  We thus transform the observed 
heliocentric radial velocities to Galactic rest frame velocities $v_{rf}$ using
	\begin{equation} v_{rf} = v_{helio} + 11.1\cos{l}\cos{b} +
247.24\sin{l}\cos{b} + 7.25\sin{b}, \label{eqn:vlsrh} \end{equation}
where $l$ and $b$ are Galactic longitude and latitude.  

	Figure \ref{fig:velh} shows the Galactic rest frame velocity distribution of
the HVS Survey stars.  Each star is drawn with its errorbars; the largest velocity
errors come from the SDSS measurements.  The lower panel shows the velocity
histogram, in 25 \kms\ bins, emphasizing the tails of the distribution.  The HVS
Survey contains mostly halo stars; the stars with $|v_{rf}|<300$ \kms\ have a 109
\kms\ line-of-sight velocity dispersion that is nearly constant with increasing
depth \citep{brown10a, gnedin10}.

	The HVS Survey also contains some remarkable velocity outliers, stars that 
are significantly unbound in radial velocity alone. The escape velocity of the Milky 
Way at 50 kpc is approximately 350 \kms\ (see Section 4.2).  We observe stars 
traveling up to +700 \kms.  Because we measure radial velocity, the full space 
motion of the stars can only be larger.

	Notably, the velocity outliers are all moving away from us, consistent with
the picture that they were ejected from the Milky Way.  The fastest star moving
towards us in the HVS Survey has $v_{rf}=-298\pm10$ \kms, consistent with Galactic
escape velocity estimates.  We re-observed the star previously reported at $-359$
\kms, SDSS J115734.45+054645.58; it is \logg$\simeq$5 white dwarf in a 13.5 hr
orbital period binary \citep{brown13a}.

	Interestingly, we observe many more bound +300 \kms\ velocity outliers
compared to $-300$ \kms\ outliers.  The orbital turn-around time for a star
traveling +300 \kms\ in the HVS Survey is about 1 Gyr.  Thus the absence of a
comparable number of $-300$ \kms\ stars demonstrates that most of the $+300$ \kms\
stars have lifetimes less than 1 Gyr \citep{brown07b, kollmeier07, yu07}.  Given
their colors (temperatures), the bound outliers in the HVS Survey are likely main
sequence B stars.

\begin{figure}          
 \plotone{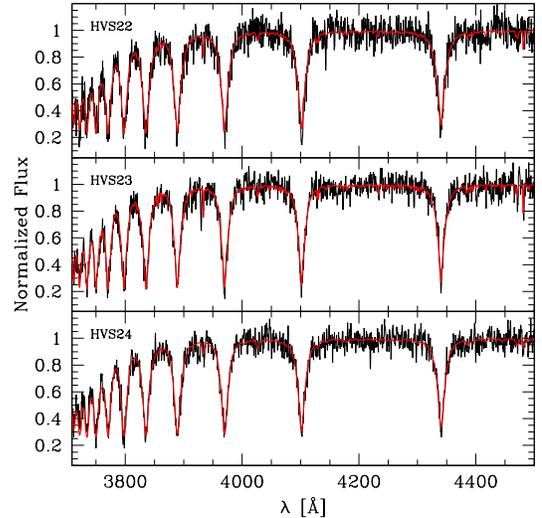}
 \caption{ \label{fig:spec22}
	HVS discovery spectra, continuum-normalized and shifted to rest-frame 
(in black), compared with the best-fitting stellar atmosphere models (red).}
 \end{figure}

\subsection{New HVSs}

	We find three HVSs in our new observations.  The star SDSS
J114146.45+044217.29, hereafter HVS22, is a faint $g=20.261\pm0.042$ and fast-moving
$v_{helio}=597.8\pm13.4$ \kms\ object.  Its minimum velocity in the Galactic rest
frame is +489 \kms.  We compare its broadband colors with stellar evolution tracks
and estimate that HVS22 is 70 kpc distant if a BHB star, and 100 kpc distant if a
main sequence star.  HVS22 is clearly unbound to the Milky Way at either distance.  
Curiously, HVS22 is located in the constellation Virgo near many of the other HVSs.  
Figure \ref{fig:spec22} shows its spectrum.

	The star SDSS J215629.02+005444.18, hereafter HVS23, is another faint 
$g=20.401\pm0.027$ star but located in the southern Galactic cap.  Its broadband 
colors imply HVS23 is 70 kpc distant if a BHB star, and 80 kpc distant if a main 
sequence star; HVS23 is unbound at either distance.  It's heliocentric velocity 
$v_{helio}=259.3\pm9.9$ \kms\ is +423 \kms\ in the Galactic rest frame.

	On the basis of follow-up observations, we re-classify SDSS
J111136.44+005856.44 as HVS24.  HVS24 has a self-consistent set of photometric and
spectroscopic distance estimates.  Its rest frame velocity $v_{rf}=+361$ \kms\ is a
significant outlier from the observed velocity distribution, and exceeds the
Galactic escape velocity at HVS24's likely distance of $R=63\pm11$ kpc.  Its rapid
rotation, described in the next Section, makes HVS24 a probable main sequence B
star. HVS24 is also located in the constellation Leo near many of the other HVSs.

\section{HVS STELLAR NATURE}

	Determining the stellar nature of HVSs is important for making accurate
distance estimates and for placing constraints on HVS ejection models.  We
investigate the nature of HVSs by performing stellar atmosphere fits to the entire
sample of HVSs.  The HVSs, unlike the other stars in the HVS Survey, were observed
multiple times to validate their radial velocities, and the summed HVS spectra have
signal-to-noise of 40--70 per resolution element adequate for fitting stellar 
atmosphere models.  \citet{heber08b} performed similar stellar atmosphere fits to 
the earliest HVS discoveries, and we now extend this analysis to the full sample of
HVSs.

	A notable result of the stellar atmosphere fits is that six of our HVSs are
fast rotators with \vsini $>170$ \kms.  Fast rotation is the unambiguous signature
of a main sequence star.  Late B-type main sequence stars have mean \vsini
$=150$~\kms\ \citep{abt02, huang06a}; evolved BHB stars have mean \vsini $=10$~\kms\
\citep{behr03}.  On this basis, we compare the HVS atmosphere parameters to
main sequence tracks to estimate masses and luminosities.

\begin{figure}          
 \plotone{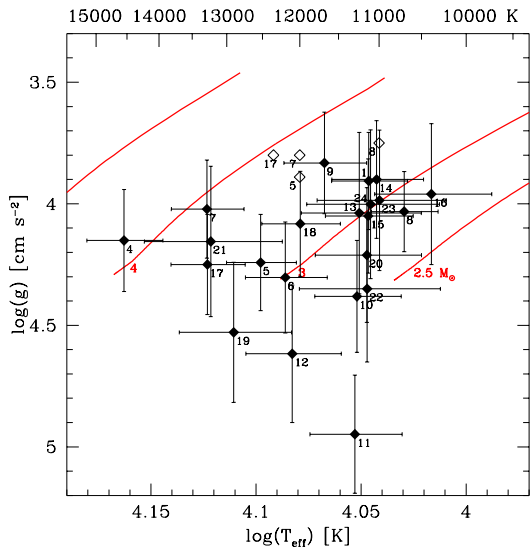}
 \caption{ \label{fig:teff}
	Derived effective temperature \teff\ and surface gravity \logg\ compared 
with Padova solar metallicity main sequence tracks for 2.5, 3, 4, and 5 \msun\ stars 
(red lines).  Parameters derived from echelle spectra for the 4 brightest HVSs (open 
diamonds) are shown for comparison. }
 \end{figure}

\subsection{Stellar Atmosphere Fits}

	Our approach to stellar atmosphere fits is identical to that described by
\citet{brown12c}.  In brief, we use ATLAS9 ODFNEW model atmosphere grids
\citep{castelli04, castelli97} to calculate synthetic spectra \citep{gray94} matched
to the spectral resolution of our observations.  We sum the rest-frame spectra of
each HVS, normalize the continuum with a low-order polynomial fit, and calculate the
$\chi^2$ of the temperature- and gravity-sensitive Balmer lines against the
synthetic models.  Finally, we fit the resulting distribution of $\chi^2$ to derive
the best-fitting parameters and uncertainties.  We find that \teff\ and \logg\ are
correlated.  A 1\% increase in \teff\ best fits a 1\% increase in \logg. Our average
uncertainties are $\pm$5\%, or $\pm$600 K, in \teff\ and $\pm$6\%, or $\pm$0.24 dex,
in \logg.

	Figure \ref{fig:spec22} shows the spectra for the new objects HVS22, HVS23,
and HVS24 compared to their best-fit stellar atmosphere model (note that we
calculate $\chi^2$ using the spectral regions around each Balmer line, and exclude
the continuum regions used for normalization).  We summarize \teff\ and \logg\
values in Table \ref{tab:hvs} and plot them in Figure \ref{fig:teff}.

	We validate our stellar atmosphere fits by comparing with previously
published results.  \citet{heber08b} perform independent stellar atmosphere fits to
our spectra of HVS1 and HVS4-7.  For this set of 5 objects, our values are on
average $400\pm400$ K hotter and $0.20\pm0.11$ dex higher in surface gravity,
consistent within the 1-$\sigma$ uncertainties but suggesting a possible systematic
offset.  Stellar atmosphere fits to high resolution echelle spectra have also been
published for the four brightest HVSs:  HVS5 \citep{brown12c}, HVS7
\citep{lopezmorales08, przybilla08b}, HVS8 \citep{lopezmorales08}, and HVS17
\citep{brown13b}.  For this small set of 4 objects, our values are on average
$600\pm400$ K hotter and $0.30\pm0.10$ dex higher in surface gravity, again
suggesting a systematic offset at the level of our 1-$\sigma$ measurement
uncertainty.  The systematic offset makes little difference to our stellar mass
estimates because the direction of the \teff--\logg\ correlation parallels the
stellar evolution tracks, but higher surface gravity causes us to systematically
under-estimate luminosity and thus distance.  The possible systematic therefore acts
as a conservative threshold on our identification of unbound HVSs.
	Given the higher precision of the echelle observations, we adopt echelle 
\teff\ and \logg\ values when available.

\begin{figure}          
 \plotone{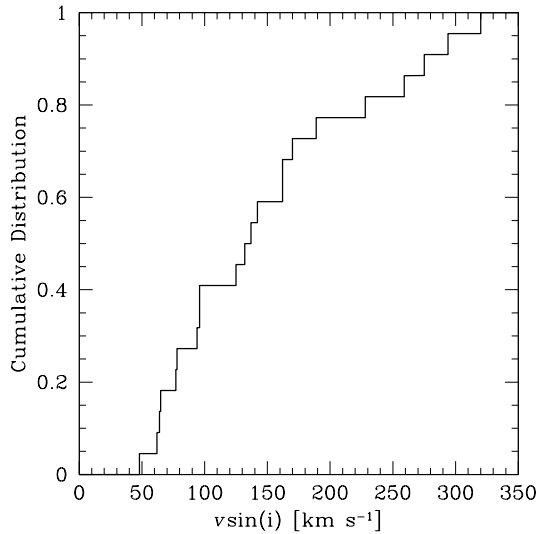}
 \caption{ \label{fig:vsini}
	Cumulative distribution of HVS projected rotation \vsini.  Six HVSs have
significant rotation \vsini$>$170 \kms, the unambiguous signature of main sequence B
stars.}
 \end{figure}

\subsection{Stellar Rotation}

	Our 1 \AA\ spectral resolution formally allows measurement of the projected 
stellar rotation for stars with \vsini$>$70 \kms.  The unresolved line Mg {\sc ii} 
$\lambda$4481, the strongest unblended metal line in the HVS spectra, in principle 
provides the best \vsini\ constraint.  In practice, our moderate spectral resolution 
and moderate S/N, combined with the nearby He {\sc i} $\lambda$4471 line, makes the 
Mg {\sc ii} $\lambda$4481 \vsini\ measurement difficult.  Instead, we rely on the 
shape of the well-sampled Balmer line profiles.  We find a clear minimum in $\chi^2$ 
when fitting model grids over \vsini, but the shallow change in $\chi^2$ implies 
large uncertainties.  We compare our \vsini\ with the four HVSs with echelle 
measurements and the independent measurement of HVS1, and find that our \vsini\ 
values are consistent within $\pm33$ \kms.  Thus, while we do not claim precise 
\vsini\ measurements, we can claim that HVSs with \vsini$>$170 \kms\ have $>$70 
\kms\ rotation at 3-$\sigma$ confidence.

	Fast rotation is interesting because it is the clear signature of a main
sequence B star.  Evolved BHB stars and main sequence B stars have very different
\vsini, as explained above.  Interestingly, the mean \vsini\ we measure for the HVS
sample is 149 \kms, equal to the expected mean for a sample of main sequence B stars
\citep{abt02, huang06a}.  This consistency is suspect, however, as the practical
lower limit of our \vsini\ measurements is about 70 \kms.  Figure \ref{fig:vsini}
plots the cumulative distribution of \vsini.

	The six HVSs with 200 to 300 \kms\ rotation in Figure \ref{fig:vsini} are
remarkable.  HVS8 is a previously known fast rotator with \vsini=260 \kms\
\citep{lopezmorales08}, but HVS9, HVS13, HVS16, HVS20, and HVS24 are new
discoveries.  In addition, HVS1, HVS6, and HVS14 have \vsini's very near 170 \kms\
\citep[see also][]{heber08b}.  Including the HVSs with echelle \vsini\ measurements,
12 (57\%) of our 21 HVSs are probable main sequence B stars on the basis of their
rotation.

\subsection{Mass and Luminosity Estimates}

	Given the observed rotation of the HVSs, we adopt main sequence stellar
evolution tracks to estimate their masses and luminosities.  Previously, we used
photometric colors and Padova tracks \citep{girardi04, marigo08} to estimate HVS
luminosities.  We now use our spectroscopic \teff\ and \logg\ with the same tracks
for consistency.  We note that the latest version of the Padova tracks adopt a
different value for solar metallicity \citep{bressan12}, but metallicity is one of
the least constrained of the HVSs' stellar parameters.  Among the objects with
echelle spectroscopy, HVS5 and HVS8 provide no metallicity constraint because of
their fast rotation.  HVS7 and HVS17 have peculiar abundance patterns and so provide
no metallicity constraint because of diffusion processes in the radiative
atmospheres of the stars.  Because these four HVSs are main sequence B stars,
however, they must have formed relatively recently in the Milky Way, presumably with
approximately solar metallicity.  Furthermore, the two HVSs with measurable iron
lines both have solar iron abundance.  We therefore adopt solar metallicity stellar
evolution tracks for estimating HVS parameters.

	Figure \ref{fig:teff} compares measured \teff\ and \logg\ to solar
metallicity Padova tracks.  The HVSs overlap the tracks for 2.5 - 4 \msun\ main
sequence stars, consistent with the underlying color selection.  HVS11 is the only
significant outlier.  Its \logg$\simeq$5 is suspiciously close to the edge of the
Kurucz model grid, but, if correct, may indicate that HVS11 is an extremely low mass
white dwarf like others found in the HVS Survey \citep{kilic07, brown13a}.  HVS11
shows no short- or long-term velocity variability, however, arguing against the low
mass white dwarf interpretation.

	We estimate stellar mass and luminosity using a Monte Carlo calculation to 
propagate the spectroscopic \teff\ and \logg\ uncertainties through the tracks.  
Thus the parameters for objects like HVS11, HVS12, and HVS19, which sit below the 
main sequence tracks, are determined by the portion of their error ellipse that 
falls on the tracks.  Table \ref{tab:hvs} summarizes the stellar mass and luminosity
estimates.

	Our spectroscopic luminosity estimates are formally no more precise than our
old photometric estimates, but they are arguably more accurate.  The mean
uncertainty of our spectroscopic absolute magnitude is $\pm$0.32 mag; the
uncertainty of the photometric absolute magnitude is $\pm$0.25 mag.  The difference
between the spectroscopic and photometric absolute magnitude $M_g$ estimates is
$0.0\pm0.39$ mag; the dispersion is consistent with the sum of the uncertainties.  
HVS1 happens to be one of the most discrepant objects:  HVS1's relatively red
$(u-g)$ color corresponds to a star with $M_{g,phot}=+0.42$, yet its hydrogen Balmer
lines correspond to a star with $M_{g,spec}=-0.35$.  We adopt the spectroscopic
estimate throughout.  Spectroscopic line profiles are immune to issues such as
photometric conditions and filter zero-point calibrations.  Thus we consider
spectroscopic measures of \teff\ and \logg\ provide the more accurate estimates of
$M_g$.

\section{HVS SPATIAL AND FLIGHT TIME DISTRIBUTIONS}

	The spatial and flight time distributions of HVSs can place useful 
constraints on their origin.  We begin by adopting an escape velocity profile for 
the Milky Way to define our sample of 21 unbound HVSs.  The unbound 2.5 - 4 \msun\ 
HVSs we observe imply there are $\simeq$100 such HVSs over the entire sky within 
$R<100$ kpc.  The ratio of HVS flight time to main sequence lifetime implies that 
the HVSs are ejected at random times during their lives.  Thus the apparent number 
of HVSs must be corrected for their finite lifetimes.

	The unbound HVSs exhibit a remarkable spatial anisotropy on the sky:  half
of the HVSs lie within a region only 15$\arcdeg$ in radius.  However, this apparent
grouping of HVSs is equally likely to share a common flight time as to be ejected
continuously.

\subsection{HVS Spatial Distribution}

\begin{figure}          
 \plotone{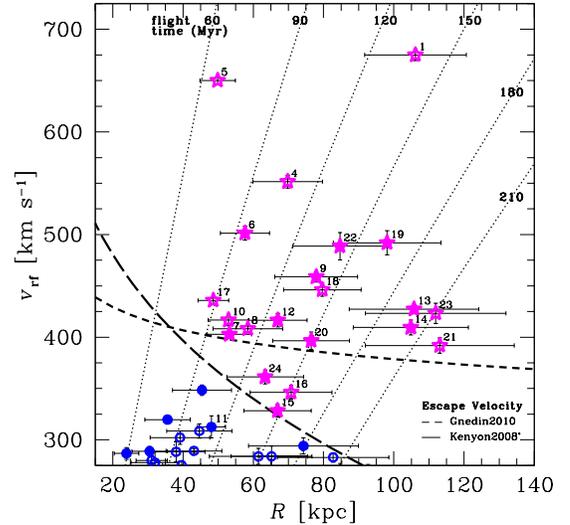}
 \caption{ \label{fig:travel}
	Galactic rest frame velocity $v_{rf}$ versus Galactocentric distance $R$.  
We adopt the escape velocity derived from the updated \citet{kenyon08} three 
component bulge-disk-halo model; we show the scaled circular velocity profile 
measured by \citet{gnedin10} for comparison.  Unbound HVSs are marked by magenta stars; 
possible bound HVSs are marked by blue circles.  Filled symbols are those objects 
clumped together around Leo.  Dotted lines are isochrones of flight time from the 
Galactic center.}
 \end{figure}

	We calculate HVS Galacto-centric radial distances, $R$, assuming the Sun is
at $R=8$ kpc.  Table \ref{tab:hvs} summarizes the results.  Figure \ref{fig:travel}
displays the Galacto-centric radial distances as a function of minimum Galactic rest
frame velocity $v_{rf}$.  Our average 32\% absolute magnitude uncertainty
corresponds to a 16\% distance uncertainty.

\begin{figure}          
 \plotone{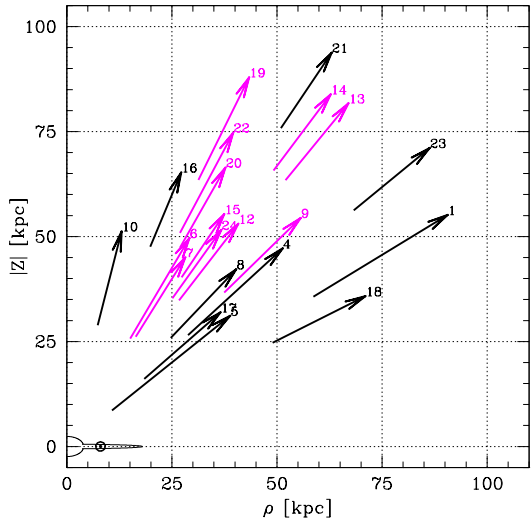}
 \caption{ \label{fig:gal}
	HVSs plotted in Galactic cylindrical coordinates.  Arrow lengths are scaled 
to $v_{rf}$, and arrow tips are located at the present positions of the HVSs.  
Magenta arrows are those objects clumped together around Leo.}
 \end{figure}

	Figure \ref{fig:gal} visualizes the spatial distribution of HVSs in Galactic
cylindrical coordinates.  The y-axis in Figure \ref{fig:gal} is the vertical
distance above the disk, the x-axis is the radial distance along the disk, and the
length of the arrows indicates the relative motion of the HVSs.  The HVSs are
presently located at the arrow tips, and span a large range of distances.  Arrows
drawn in magenta are the HVSs located in the clump around the constellation Leo. To
properly establish our sample of unbound HVSs we must define the escape velocity of
the Milky Way.

\subsection{Galactic Escape Velocity}

	The escape velocity of the Milky Way varies with distance because of the
Galaxy's extended mass distribution.  We estimate the Milky Way's escape velocity
using an updated version of the \citet{kenyon08} three component bulge-disk-halo
potential model that fits observed mass measurements.  We adopt a new disk mass
$M_d=6\times10^{10}$ \msun\ and radial scale length $a_d=2.75$ kpc that yields a
flat 235 \kms\ rotation curve, consistent with our adopted circular velocity.  We
leave the other potential model parameters unchanged.  Compared to the original
\citet{kenyon08} model, our updated model contains more disk mass and thus requires
a slightly larger ejection velocity to escape.

	Escape velocity is not formally defined in the three-component potential
model.  We empirically estimate escape velocity by dropping a test particle from the
virial radius at 250 kpc.  The resulting escape velocity curve drawn in Figure
\ref{fig:travel} can be fit with the following relation, valid in the range
$15<R<150$ kpc:
	\begin{eqnarray} \nonumber v_{esc}(R) &= 624.9 -9.4136 R +0.134197 R^2 
-1.28165\times10^{-3} R^3 \\ &+6.47686\times10^{-6} R^4 -1.31814\times10^{-8} R^5.
	\end{eqnarray} Our adopted potential model yields an escape velocity of
366 \kms\ at $R=50$ kpc and 570 \kms\ at $R=8$ kpc.  The latter value is consistent
with current solar neighborhood escape velocity measurements \citep{smith07,
piffl13}.

	An alternative escape velocity estimate is provided by scaling the halo
circular velocity measured by \citet{gnedin10} from a Jeans analysis of the Milky
Wau stellar halo velocity dispersion profile.  The inferred escape velocity in this
case is 400 \kms\ at $R=50$ kpc (see Figure \ref{fig:travel}).  Scaling circular
velocity to estimate escape velocity is formally incorrect, however.  More recently,
\citet{rashkov13} argue that the Milky Way halo is significantly less massive than
assumed by \citet{kenyon08} and measured by \citet{gnedin10}.  Given the
uncertainties, we consider the updated \citet{kenyon08} escape velocity model 
a reasonable choice.

	Given the above definition of escape velocity, we identify 21 HVSs that are
unbound on the basis of radial velocity alone (Figure \ref{fig:travel}).  We include
HVS16 and HVS24 because they sit well above the \citet{kenyon08} escape velocity
curve and have \vsini$>$200 \kms.  Thus HVS16 and HVS24 are $\simeq$3 \msun\ B stars
at 60--70 kpc distances.  HVS15 is a borderline case that we consider a likely HVS
because of its significant $v_{rf}=328$ \kms\ velocity and its self-consistent
photometric and spectroscopic distance estimates.

	HVS11, given our new distance estimate, is an object that we now classify as
a possible ``bound HVS.'' In fact, we consider all of the remaining objects with 275
\kms\ $<v_{rf}<v_{esc}$ as possible bound HVSs:  significant velocity outliers that
are bound.  Our choice of 275 \kms\ is motivated by the relative absence of stars
with velocities less than $-275$ \kms\ in our Survey.  We emphasize that our choice
of the 275 \kms\ threshold is appropriate only in the context of our halo radial
velocity survey, and is {\it not} a generalizable threshold on which to select HVSs
in other contexts, such as the solar neighborhood or the Galactic center
\citep[e.g.][]{zubovas13} where escape velocity is very different.

\begin{figure}          
 \plotone{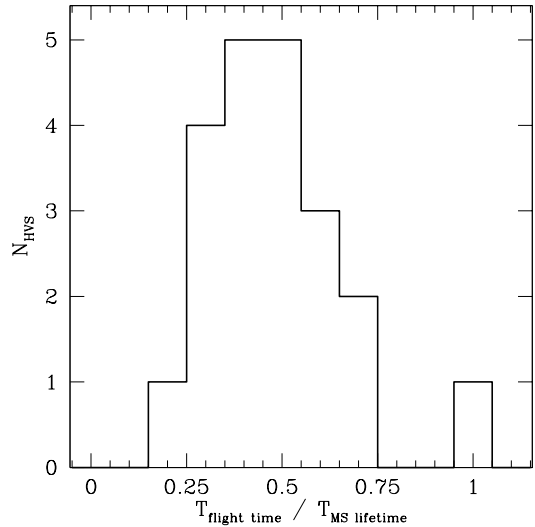}
 \caption{ \label{fig:lifetime}
	Ratios of HVS flight time to main sequence lifetime.  An average ratio of
0.5 suggests that HVSs are ejected at random times during their main sequence
lifetimes.}
 \end{figure}

\begin{figure*}          
 \plottwo{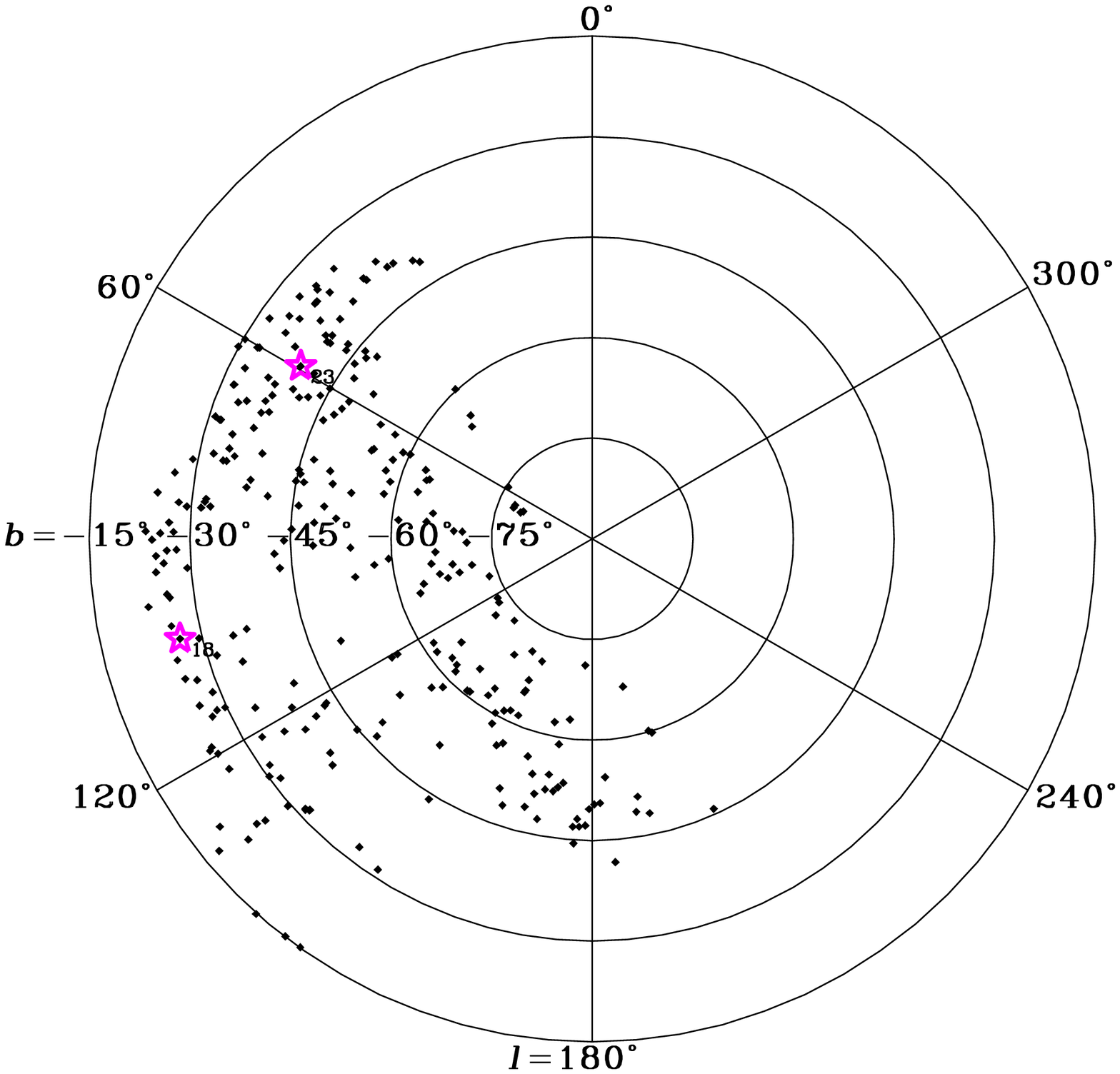}{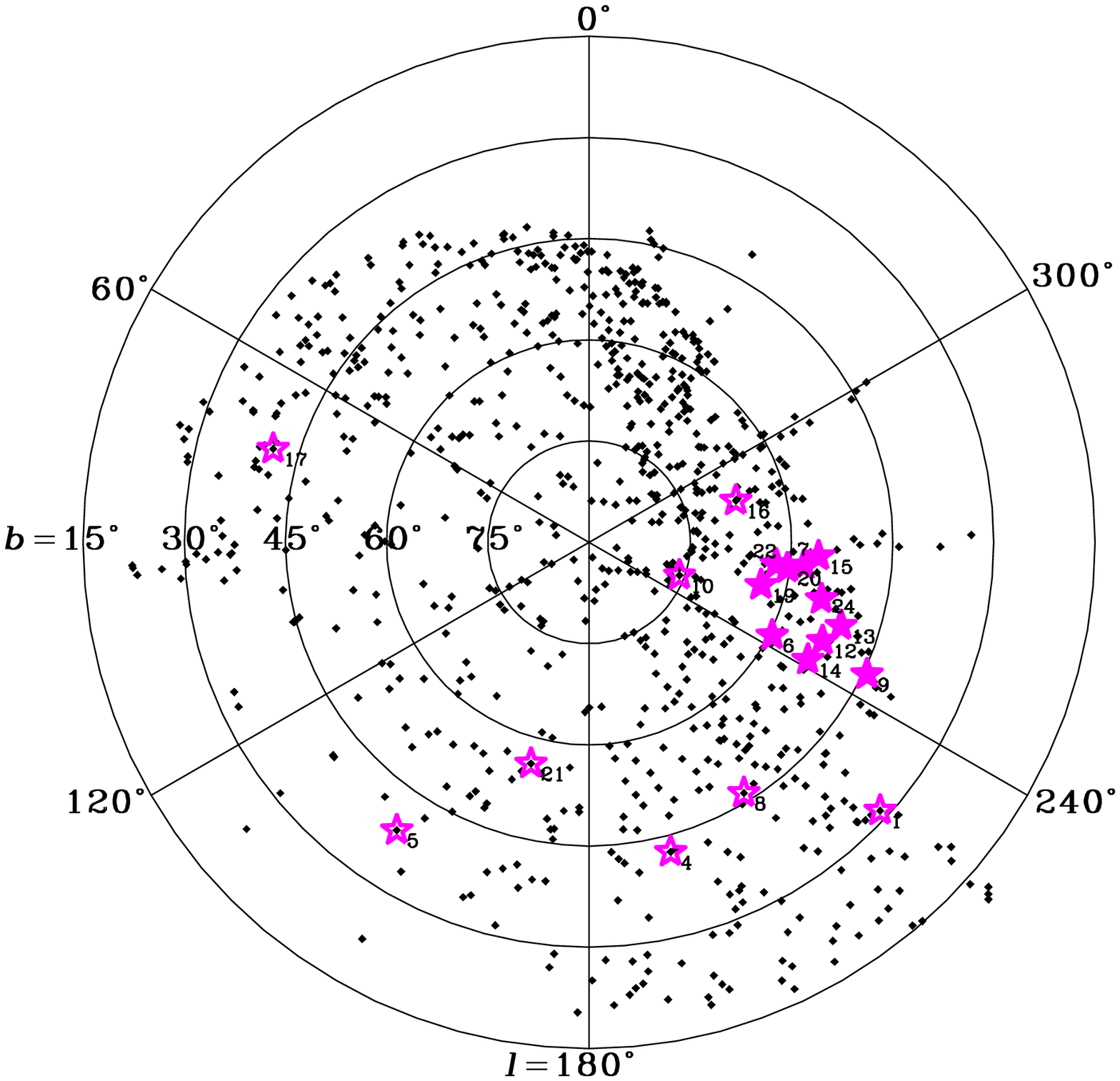}
 \caption{ \label{fig:polar}
	Polar projections, in Galactic coordinates, showing the spatial distribution 
of the 1126 late B-type stars in the HVS Survey.  Southern Galactic cap is left and 
the northern Galactic cap is right.  The 21 unbound HVSs are marked by magenta 
stars; filled stars mark the HVSs clumped together around Leo. }
 \end{figure*}

\subsection{HVS Flight Times and Main Sequence Lifetimes}

	We now use our distance and velocity measurements to estimate HVS flight
times.  Our approach is to start at the present location and velocity of each HVS,
and then calculate its trajectory backwards through the Galactic center in the
updated \citet{kenyon08} potential model (see the flight time isochrones in Figure
\ref{fig:travel}).  Our flight times are formally upper limits because we assume
that the observed radial velocities are the full space motion of the HVSs.  This
assumption is reasonable for objects on radial trajectories at large 50 - 120 kpc
distances.  We note that flight times from other locations in the Milky Way, such as
the solar circle $R=8$ kpc, typically vary by only $\pm$3 Myr ($\sim$2\%) from the
Galactic center flight time because of the large distances and high Galactic
latitudes of the HVSs.  We estimate flight time uncertainties by propagating the
velocity and distance errors through our trajectory calculations.  Distance errors
dominate the uncertainty; thus our typical flight time uncertainties are 16\%, or
about $\pm$22 Myr.

	The difference between flight time and age can constrain a HVS's origin
\citep{brown12c}, however our spectroscopic measurements provide no constraint on
HVS age.  The best we can do is to use our stellar mass estimates to estimate a
star's total main sequence lifetime.  In the Padova tracks, the main sequence
lifetime of a 3 \msun\ star is 350 Myr and a 4 \msun\ star is 180 Myr.

	Figure \ref{fig:lifetime} compares HVS flight times and main sequence
lifetimes.  In all cases, the ratios of HVS flight time to main sequence lifetime
are less than one.  In other words, the HVSs are all consistent with being main
sequence stars ejected from the Milky Way.  Interestingly, the average ratio of HVS
flight time to main sequence lifetime is 0.46.  A ratio of one-half suggests that
HVSs are ejected at random times during their lives.

\subsection{HVS Spatial Anisotropy}

	The spatial distribution of HVSs is interesting because it is linked to
their origin.  HVSs can in principle appear anywhere in our 12,000 deg$^2$ survey
because the central MBH can in principle eject a HVS in any direction.  Yet eleven
(52\%) of the 21 unbound HVSs are located in a $25\arcdeg\times25\arcdeg$ (5\% of
Survey area) region at the edge of our survey, centered around (RA, Dec) =
($11^h10^m00^s, +3^d00^m00^s$) J2000 in the direction of the constellation Leo.

	Figure \ref{fig:polar} plots the spatial distribution of every star observed
in the HVS Survey in two polar projections centered on the north and south Galactic
poles.  The overall distribution of stars reflects the SDSS imaging footprint.  In
this footprint, our Survey stars have an approximately isotropic distribution,
although part of the Sgr dwarf galaxy tidal stream is visible as an overdensity of
stars arcing to the right of the north Galactic pole \citep{king12}.  The unbound
HVSs are marked by magenta stars in Figure \ref{fig:polar}.  The new HVS discoveries
strengthen the previously claimed HVS spatial anisotropy.  By any measure, whether
Galactic longitude distribution, angular separation distribution, or the two-point
angular correlation function, unbound HVSs are significantly clustered
\citep[see][]{brown09b, brown12b}.  The lower velocity, possible bound HVSs have a
more isotropic distribution.

	Various models have been proposed to explain the HVS anisotropy.  Each model 
predicts different spatial and flight time distribution of HVSs.  One model is the 
in-spiral of two massive black holes in the Galactic center \citep{gualandris05, 
levin06, sesana06, baumgardt06}.  A binary black hole preferentially ejects HVSs 
from its orbital plane; thus HVSs ejected by a binary black hole in-spiral event 
should form a ring around the sky.  Models predict a binary black hole will harden 
and merge on $\sim$1 Myr timescales \citep{sesana08}.  Thus the signature of a 
single in-spiral event is a ring of HVSs with a common flight time.

	\citet{abadi09} propose that the HVS anisotropy comes from the stellar
ejecta of a tidally disrupted dwarf galaxy \citep[see also][]{piffl11}.  This model
predicts a single clump of HVS with a common flight time.  However, it is unclear
where the progenitor would come from.  No Local Group dwarf galaxy has a velocity
comparable to the HVSs, and a gas-rich star forming galaxy is required to explain
the B-type HVSs.  There are no unbound A- or F-type stars in same region of sky
\citep{kollmeier09, kollmeier10} as one expects for a disrupted dwarf galaxy.

	\citet{lu10} and \citet{zhang10, zhang13} propose that the HVS anisotropy
reflects the anisotropic distribution of stars in the Galactic center.  If HVSs are
ejected by the central MBH, then the direction of ejection corresponds to the
direction that their progenitors encounter the MBH.  Interestingly, known HVSs fall
on the projected orbital planes of the clockwise and counter-clockwise disks of
stars that presently orbit Sgr A*.  There is no explanation for the confinement of
HVS ejections to two fixed planes over the past 200 Myr, however.  If we accept
fixed ejection planes, this model allows for bands of HVSs on the sky containing
stars of all possible flight times.

	Finally, \citet{brown12b} propose that the HVS anisotropy may reflect the
anisotropy of the underlying Galactic gravitational potential.  Stars ejected along
the long axis of the potential are decelerated less than those ejected along the
minor axis.  An initially isotropic distribution of marginally unbound HVSs can thus
appear anisotropic in the halo.  The predicted distribution of HVSs depends on the
axis ratio and the rotation direction of the potential.  If there is rotation around
the long axis of the potential, for example, HVSs should appear in two clumps on
opposite sides of the sky with all possible flight times.

	Although the present spatial distribution of HVSs can be described by two
planes \citep{lu10, zhang10}, this is not necessarily the required distribution.  
The observed clump of HVSs abuts one edge of the Survey -- the edge defined by the
celestial equator.  We require a southern hemisphere HVS Survey to see the full
all-sky distribution.  For now, we simply test whether the observed clump of HVSs
share a common flight time.

\begin{figure}          
 \plotone{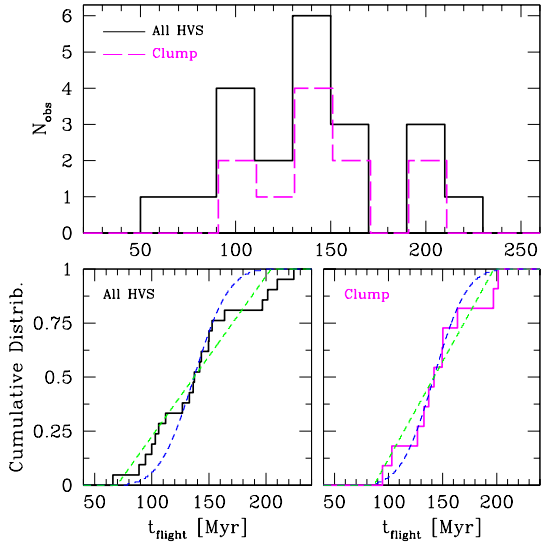}
 \caption{ \label{fig:distt}
	Distribution of HVS flight times from the Galactic center, in 20 Myr bins 
(upper panel).  The subset of HVSs clumped together around Leo are drawn in magenta.  
Lower panels plot the cumulative distributions of the full sample (left) and the 
clumped sample (right) compared to a continuous distribution and a Gaussian 
distribution with a 24 Myr dispersion. }
 \end{figure}

\subsection{Flight Time Constraints on Origin}

	HVS flight times provide another constraint on origin.  Figure
\ref{fig:distt} plots the distribution of flight times for the full sample of HVSs
and for the clump of HVSs within 15$\arcdeg$ of $11^h10^m00^s, +3^d00^m00^s$
(J2000).  We make this particular selection because it cuts the HVS sample in half,
and allows us to test the dwarf galaxy tidal debris hypothesis.  We calculate the
non-parametric Kolmogorov-Smirnov statistic of the samples against two models.  The
first model is a constant distribution, normalized to the number of HVSs and
centered on the mean $t_{flight}$ of each HVS sample.  The second model is a burst
distribution: a Gaussian with identical mean and normalization as the first model,
but a dispersion equal to 16\% of the mean $t_{flight}$ appropriate for our
measurement uncertainties.  The lower panels of Figure \ref{fig:distt} compare these
models to the observed cumulative distribution of HVS flight times.

	The full HVS sample is well-described by a continuous flight time
distribution (KS probability 0.80) but poorly described by a single burst (KS
probability 0.08).  The spatially clumped HVS flight times, on the other hand, are
equally well-described by a continuous distribution (KS probability 0.89) and a
single burst (KS probability 0.83).  Thus we can neither confirm nor deny the tidal
debris origin on the basis of the spatially clumped HVS flight times.  On the other
hand, the flight times of the full set of HVSs rule out a single burst.  In other
words, if HVSs are ejected by binary MBH in-spiral or dwarf galaxy tidal disruption
events, the data require multiple events to explain the full HVS sample.

\section{HVS PROPER MOTION PREDICTIONS}

	Proper motions may one day provide a direct constraint on HVS origin.  The
proper motions of the HVSs remain unmeasured except in special cases
\citep{brown10b}, because the HVSs are distant and their proper motions are too
small $\lesssim$1 \mas\ to be measured in ground-based catalogs.  {\it Gaia}
promises to measure proper motions for the HVSs with 0.1 \mas\ precision.
These measurements will, in many cases, discriminate between Galactic center and 
disk origins.  

	Here, we lay the groundwork for interpreting future proper motion
measurements by 1) predicting the proper motion HVSs must have if they come from the
Galactic center, and 2) calculating the ejection velocities required for alternative
Galactic disk origins.  Similar calculations were done for early HVS discoveries
\citep{svensson08}.  Here we address the full sample of unbound HVSs.

	Our computational approach is to step through a grid of all possible proper
motions and calculate the corresponding HVS trajectories backwards in time through
the updated \citet{kenyon08} potential model.  For each trajectory we record
Galactic plane-crossing location and velocity, and we calculate the necessary
Galactic disk ejection velocity assuming a flat 235 \kms\ rotation curve.

\begin{figure*}          
 \plotone{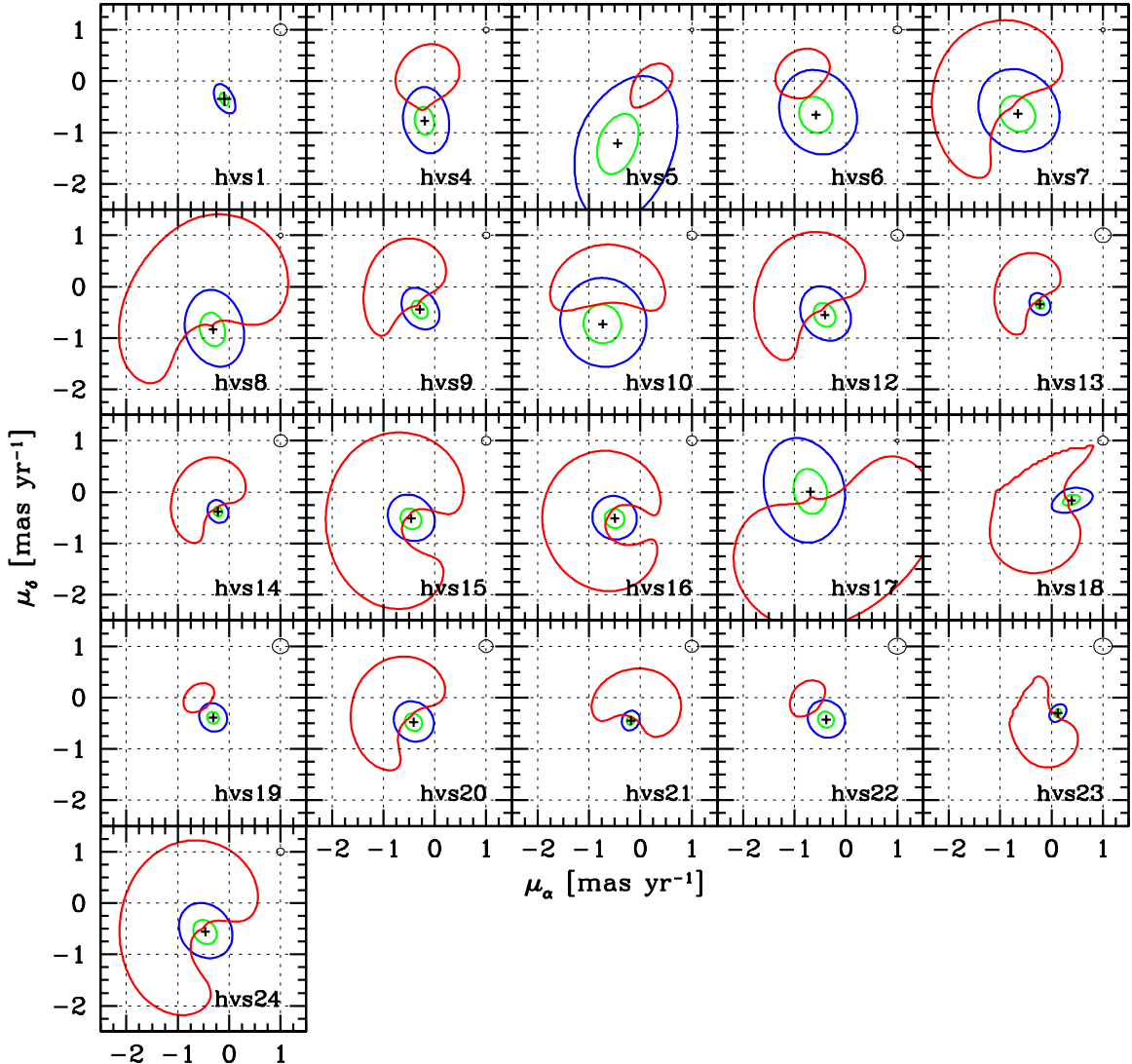}
 \caption{ \label{fig:pm5}
	HVS trajectories mapped into proper motion space.  Green and blue ellipses
are the locus of trajectories that pass within 8 and 20 kpc, respectively, of the
Galactic center.  Trajectories that pass through the Galactic center are marked by
an +.  The locus of trajectories with disk plane ejection velocities of 600 \kms,
the escape velocity from the surface of a 3 \msun star, are indicated by the red
contours.  Physically allowed disk ejections are mostly limited to the outer disk.  
Black circles are the predicted {\it Gaia} proper motion error ellipses for each 
HVS. }
 \end{figure*}

	Figure \ref{fig:pm5} shows how the trajectories for each HVS map onto proper
motion space:  the blue and green ellipses are the locus of proper motions with
trajectories crossing the Galactic plane at $R=20$ kpc and $R=8$ kpc, respectively.  
We identify a Galactic center origin by the trajectory with the smallest Galactic
pericenter passage.  An x marks this trajectory, and Table \ref{tab:hvs} lists the 
corresponding proper motion for the Galactic center origin.

	Our trajectory calculations place interesting constraints on a disk origin.  
Theorists show that the maximum possible ejection velocity from stellar
binary-binary interactions is, under a point mass assumption, the escape velocity
from the surface of the most massive star \citep{leonard91}.  The escape velocity
from the surface of a 3 \msun\ main sequence star is $\simeq$600 \kms.  Thus, the
only locations where HVSs can be ejected by stellar runaway mechanisms are those
locations in the disk where the ejection velocity is below 600 \kms.  The red
contours in Figure \ref{fig:pm5} shows this constraint for each HVS.  Trajectories
with disk ejection velocities $<$600 \kms\ have proper motions that lie inside the
red contour.

	Interestingly, there is {\it no} location where HVS1 can physically be
ejected from the Galactic disk.  Due to its extreme velocity and distance, HVS1
provides a strong case for a Galactic center origin.  For the other HVSs, there are
finite but limited portions of the disk from where they can be ejected.  The lowest
disk ejection velocities occur where disk rotation points towards the HVSs.  
Minimum disk ejection velocities range 400--590 \kms.  The average radial location
of the minima is $R=28$ kpc, however, well outside the Milky Way's stellar disk.  
Only HVS5, with a disk ejection velocity of 590 \kms\ at $R=18$ kpc, has a minimum
that falls within the $R\simeq20$ kpc extent of the observed stellar disk.  The
viable region for HVS disk ejections is thus typically a small fraction of the disk;  
it is the area included within both the blue and red contours in Figure
\ref{fig:pm5}.

	Notably, a Galactic center origin is often well-separated in proper motion
from a disk origin.  This separation occurs because viable disk ejection
trajectories are from the outer disk.  Future proper motion measurements with 0.1
\mas\ precision should thus be able to distinguish between Galactic center and disk
origins.  {\it Gaia} astrometric performance specifications predict 0.035 - 0.17
\mas\ proper motion uncertainties for the
HVSs\footnote{\url{http://www.cosmos.esa.int/web/gaia/science-performance}}; the
error ellipses depend on apparent magnitude and are drawn as the black circles in
the upper right-hand corners of the panels in Figure \ref{fig:pm5}.  If {\it Gaia}
meets its predicted astrometric performance, it should unambiguously determine the
origin of HVS4, HVS5, HVS6, HVS7, HVS8, HVS9, HVS10, and HVS17.

\section{DISCUSSION AND CONCLUSIONS}

	The HVS Survey is a complete, color-selected sample of late B-type stars
over 12,000 deg$^2$ (29\%) of sky.  Spectroscopy of these stars reveals 21 HVSs
unbound in radial velocity alone.  The Survey also identifies a comparable number of
velocity outliers that are possibly bound HVSs.  Stellar atmosphere fits show that,
on the basis of projected stellar rotation, at least half of the HVSs are certain
main sequence 2.5 - 4 \msun\ stars at 50 - 120 kpc distances.

	If we assume that the HVSs are ejected continuously and isotropically, the
21 observed HVSs implies there are $\simeq$100 unbound 2.5 - 4 \msun\ HVSs over the
entire sky within $R<100$ kpc.  This calculation neglects HVSs missing from our
Survey because of their short lifetimes, however.  In previous papers, we corrected
the observed number of HVSs by the fraction that do not survive to reach large
distances assuming the HVSs are ejected at zero age \citep{bromley06, brown07b}.  
This assumption is flawed, as demonstrated by the observed ratios of HVS flight time
to main sequence lifetime.  If we instead assume that HVSs are ejected at any time
during their lifetime, then we expect about 32\% of 2.5 - 4 \msun\ HVSs do not
survive to reach 50 kpc, and 64\% do not survive to reach 100 kpc.  The corrected
number of HVSs is $\simeq$300 unbound 2.5 - 4 \msun\ HVSs over the entire sky within
$R<100$ kpc.  The ejection rate of unbound 2.5 - 4 \msun\ HVSs is thus
$1.5\times10^{-6}$ yr$^{-1}$.

	To infer a total HVS ejection rate requires many more assumptions. The 
simplest approach is to assume that all stellar masses have identical binary 
fractions, binary orbital distributions, and ejection velocities, and then scale the 
observed 2.5 - 4 \msun\ HVSs by an assumed mass function.  For a Salpeter IMF, 
integrated over the mass range $0.1 < M < 100$ \msun\ and scaled to the corrected 
number of 2.5 - 4 \msun\ HVSs, the total HVS ejection rate is $2.5\times10^{-4}$ 
yr$^{-1}$.  This rate is comparable to the $10^{-4}$ yr$^{-1}$ HVS ejection rates 
predicted by theory \citep{hills88, perets07, zhang13}.  It is difficult to take the 
ejection rate calculation further because the properties of binary stars and the 
stellar mass function in the Galactic center are poorly constrained.  However, the 
general agreement between HVS observation and HVS theory is a good consistently 
check.

	The distribution of HVS flight times provides another constraint on origin.  
HVSs ejected by a single MBH can be observed with any flight time.  HVSs ejected by
a dwarf galaxy tidal disruption event or a binary MBH in-spiral event must share a
common flight time.  Our full HVS sample is well-described by a continuous flight
time distribution, but poorly described by a single burst.  This conclusion is
consistent with the \citet{hills88} scenario.  Alternatively, this conclusion
requires multiple binary MBH in-spiral or galaxy tidal disruption events.

	An intriguing result remains the anisotropic HVS spatial distribution:  
half of the observed HVSs are located around the constellation Leo.  Because the
HVSs are clumped at the edge of our Survey, the true distribution of HVSs remains
unknown.  The flight times of the spatially clumped HVSs also provide no constraint;
the flight times of the clump are equally well-described by a continuous
distribution or a single burst.

	In the near future, proper motions promise to provide a direct test of HVS
origin.  The first unbound main sequence star with measured proper motion is HD
271791 \citep{heber08}, a runaway B star ejected in the direction of rotation from
the outer disk by a stellar binary disruption process \citep{przybilla08c,
gvaramadze09}.  Stellar binary disruption processes have a speed limit, however, set
by the escape velocity from the surface of the star.  We perform trajectory
calculations for our HVSs and show that, in many cases, physically allowed
disk-ejection trajectories differ systematically from Galactic center trajectories
in proper motion space.  Proper motions with 0.1 \mas\ precision, possible with
{\it Hubble Space Telescope} and {\it Gaia}, should place clear constraints on the
origin of HVSs 4 - 10 and HVS17.

	We also look forward to performing a southern hemisphere HVS Survey in the
near future.  Whether the HVSs are ejected in rings, clumps, or a more isotropic
distribution is an important constraint on their origin.  Doubling the sample of
HVSs to $\sim$50 objects should also allow us to discriminate between single and
binary MBH ejection mechanisms on the basis of the HVS velocity distribution
\citep{sesana07b, perets09a}.  Ejection models predict $>$1,000 \kms\ stars
\citep[e.g.][]{bromley06, sesana07b, zhang10, rossi14}, stars that are not yet
observed.  We expect that full sky coverage provided by surveys like SkyMapper
\citep{keller07} will provide a rich source of future HVS discoveries.

\acknowledgements

	We thank A.\ Milone, J. McAfee, S.\ Gottilla, and E. Martin for their
assistance with observations obtained at the MMT Observatory, a joint facility of
the Smithsonian Institution and the University of Arizona.  This project makes use
of data products from the Sloan Digital Sky Survey, which is managed by the
Astrophysical Research Consortium for the Participating Institutions.  This research
makes use of NASA's Astrophysics Data System Bibliographic Services.  This work was
supported by the Smithsonian Institution.

{\it Facilities:} \facility{MMT (Blue Channel Spectrograph)}

\begin{deluxetable*}{ccccccccccccc}
\tabletypesize{\scriptsize}
\tablecaption{HVS Survey Stars with $v_{rf}>+275$ \kms\label{tab:hvs}}
\tablewidth{0pt}
\tablecolumns{13}
\tablehead{
  \colhead{HVS} & \colhead{$v_{\sun}$} & \colhead{$v_{rf}$} &
  \colhead{$g_0$} & \colhead{\teff} & \colhead{\logg} & \colhead{\vsini} &
  \colhead{mass} & \colhead{$M_g$} & \colhead{$R_{GC}$} &
  \colhead{$t_{flight}$} & \colhead{$(\mu_\alpha ,~ \mu_\delta)_{GC}$} & 
  \colhead{Catalog} \\
  \colhead{} & \colhead{(\kms )} & \colhead{(\kms )} &
  \colhead{(mag)} & \colhead{(K)} & \colhead{(cm s$^{-2}$)} & \colhead{(\kms )} &
  \colhead{(\msun )} & \colhead{(mag)} & \colhead{(kpc)} &
  \colhead{(Myr)} & \colhead{(\mas)} & \colhead{}
}
        \startdata
\cutinhead{HVSs}
 1 & $833.0\pm 5.5$ & 675.0 & 19.688 & $11125\pm 463$ & $3.91\pm0.20$ & 162 & $3.21\pm0.13$ & $-0.34\pm0.31$ & $106\pm14$ & $136\pm19$ & $-0.09,-0.35$ & SDSS J090744.99$+$024506.89 \\
 4 & $600.9\pm 6.2$ & 551.5 & 18.314 & $14547\pm 607$ & $4.15\pm0.21$ &  77 & $4.24\pm0.16$ & $-0.71\pm0.34$ & $ 70\pm10$ & $106\pm17$ & $-0.20,-0.77$ & SDSS J091301.01$+$305119.83 \\
 5 & $545.5\pm 4.3$ & 650.1 & 17.557 & $12000\pm 350$ & $3.89\pm0.13$ & 132 & $3.58\pm0.11$ & $-0.68\pm0.25$ & $ 50\pm 5$ & $ 66\pm 7$ & $-0.44,-1.21$ & SDSS J091759.47$+$672238.35 \\
 6 & $609.4\pm 6.8$ & 501.4 & 18.966 & $12190\pm 546$ & $4.30\pm0.23$ & 170 & $3.06\pm0.11$ & $+0.25\pm0.28$ & $ 58\pm 7$ & $ 94\pm13$ & $-0.58,-0.66$ & SDSS J110557.45$+$093439.47 \\
 7 & $527.8\pm 2.7$ & 402.8 & 17.637 & $12000\pm 500$ & $3.80\pm0.10$ &  62 & $3.76\pm0.13$ & $-0.95\pm0.27$ & $ 53\pm 6$ & $103\pm12$ & $-0.65,-0.63$ & SDSS J113312.12$+$010824.87 \\
 8 & $499.3\pm 2.9$ & 408.3 & 17.939 & $11000\pm1000$ & $3.75\pm0.25$ & 320 & $3.42\pm0.20$ & $-0.70\pm0.40$ & $ 58\pm10$ & $112\pm19$ & $-0.31,-0.84$ & SDSS J094214.03$+$200322.07 \\
 9 & $616.8\pm 5.1$ & 458.8 & 18.639 & $11680\pm 529$ & $3.83\pm0.21$ & 294 & $3.56\pm0.16$ & $-0.74\pm0.34$ & $ 78\pm12$ & $137\pm21$ & $-0.29,-0.44$ & SDSS J102137.08$-$005234.77 \\
10 & $467.9\pm 5.6$ & 416.7 & 19.220 & $11270\pm 533$ & $4.38\pm0.23$ &  64 & $2.65\pm0.11$ & $+0.65\pm0.24$ & $ 53\pm 6$ & $100\pm12$ & $-0.73,-0.73$ & SDSS J120337.85$+$180250.35 \\
12 & $552.2\pm 6.6$ & 416.5 & 19.609 & $12098\pm 632$ & $4.62\pm0.28$ &  78 & $2.73\pm0.14$ & $+0.56\pm0.28$ & $ 67\pm 8$ & $127\pm17$ & $-0.41,-0.55$ & SDSS J105009.59$+$031550.67 \\
13 & $572.7\pm 4.5$ & 427.3 & 20.018 & $11241\pm 739$ & $4.04\pm0.33$ & 189 & $3.07\pm0.17$ & $-0.07\pm0.39$ & $106\pm18$ & $197\pm37$ & $-0.23,-0.34$ & SDSS J105248.30$-$000133.94 \\
14 & $537.3\pm 7.2$ & 409.4 & 19.717 & $11030\pm 554$ & $3.90\pm0.24$ & 162 & $3.18\pm0.15$ & $-0.33\pm0.35$ & $105\pm16$ & $202\pm32$ & $-0.22,-0.38$ & SDSS J104401.75$+$061139.02 \\
15 & $461.0\pm 6.3$ & 328.3 & 19.153 & $11132\pm 535$ & $4.05\pm0.23$ & 125 & $2.99\pm0.12$ & $+0.05\pm0.32$ & $ 67\pm10$ & $150\pm23$ & $-0.46,-0.52$ & SDSS J113341.09$-$012114.25 \\
16 & $429.8\pm 7.0$ & 346.2 & 19.334 & $10388\pm 666$ & $3.96\pm0.29$ & 259 & $2.85\pm0.15$ & $+0.07\pm0.36$ & $ 71\pm12$ & $153\pm26$ & $-0.50,-0.51$ & SDSS J122523.40$+$052233.84 \\
17 & $250.2\pm 2.9$ & 435.8 & 17.427 & $12350\pm 290$ & $3.80\pm0.09$ &  96 & $3.91\pm0.09$ & $-1.05\pm0.19$ & $ 49\pm 4$ & $ 89\pm 8$ & $-0.69,+0.01$ & SDSS J164156.39$+$472346.12 \\
18 & $237.3\pm 6.4$ & 446.2 & 19.302 & $11993\pm 516$ & $4.08\pm0.22$ &  96 & $3.27\pm0.12$ & $-0.15\pm0.31$ & $ 80\pm11$ & $143\pm22$ & $+0.39,-0.16$ & SDSS J232904.94$+$330011.47 \\
19 & $592.8\pm11.8$ & 492.0 & 20.061 & $12900\pm 793$ & $4.53\pm0.29$ & 137 & $3.12\pm0.17$ & $+0.13\pm0.34$ & $ 98\pm15$ & $164\pm28$ & $-0.31,-0.39$ & SDSS J113517.75$+$080201.49 \\
20 & $512.1\pm 8.5$ & 396.6 & 19.807 & $11149\pm 649$ & $4.21\pm0.28$ & 275 & $2.79\pm0.12$ & $+0.42\pm0.32$ & $ 76\pm11$ & $150\pm24$ & $-0.41,-0.48$ & SDSS J113637.13$+$033106.84 \\
21 & $356.8\pm 7.5$ & 391.9 & 19.730 & $13229\pm 998$ & $4.16\pm0.31$ &  65 & $3.70\pm0.21$ & $-0.45\pm0.42$ & $113\pm21$ & $224\pm48$ & $-0.18,-0.45$ & SDSS J103418.25$+$481134.57 \\
22 & $597.8\pm13.4$ & 488.7 & 20.181 & $11145\pm 859$ & $4.35\pm0.30$ &  94 & $2.66\pm0.15$ & $+0.56\pm0.35$ & $ 85\pm13$ & $142\pm25$ & $-0.38,-0.43$ & SDSS J114146.44$+$044217.29 \\
23 & $259.3\pm 9.8$ & 423.2 & 20.201 & $10996\pm 778$ & $3.99\pm0.29$ &  48 & $3.04\pm0.16$ & $-0.10\pm0.38$ & $112\pm20$ & $210\pm40$ & $+0.12,-0.30$ & SDSS J215629.01$+$005444.18 \\
24 & $496.2\pm 6.8$ & 361.3 & 18.855 & $11103\pm 806$ & $4.00\pm0.31$ & 228 & $3.06\pm0.17$ & $-0.10\pm0.39$ & $ 63\pm11$ & $133\pm25$ & $-0.46,-0.56$ & SDSS J111136.44$+$005856.44 \\
\cutinhead{Possible Bound HVSs}
   & $148.0\pm 6.9$ & 308.6 & 17.139 & $12383\pm1474$ & $3.81\pm0.51$ & 258 & $3.90\pm0.32$ & $-0.94\pm0.49$ & $ 45\pm 9$ & $104\pm22$ & $+0.93,-0.54$ & SDSS J002810.33$+$215809.66 \\
   & $138.3\pm 6.5$ & 302.1 & 17.767 & $12403\pm1632$ & $4.54\pm0.60$ & 328 & $2.99\pm0.33$ & $+0.07\pm0.54$ & $ 39\pm 9$ & $ 92\pm22$ & $+1.22,-0.55$ & SDSS J005956.06$+$313439.29 \\
   & $352.7\pm 4.5$ & 282.8 & 18.388 & $10613\pm 908$ & $3.52\pm0.44$ & 220 & $3.61\pm0.25$ & $-1.00\pm0.45$ & $ 83\pm16$ & $203\pm36$ & $+0.09,-0.64$ & SDSS J074950.24$+$243841.16 \\
   & $216.1\pm 4.4$ & 279.9 & 17.276 & $12027\pm 880$ & $4.50\pm0.37$ & 176 & $2.85\pm0.19$ & $+0.34\pm0.39$ & $ 31\pm 4$ & $ 77\pm12$ & $-0.19,-2.22$ & SDSS J081828.07$+$570922.07 \\
   & $298.6\pm 4.3$ & 275.2 & 18.081 & $11673\pm 798$ & $4.46\pm0.33$ &  45 & $2.74\pm0.16$ & $+0.47\pm0.35$ & $ 40\pm 5$ & $ 99\pm15$ & $-0.33,-1.54$ & SDSS J090710.07$+$365957.54 \\
11 & $472.7\pm10.8$ & 312.6 & 19.582 & $11296\pm 571$ & $4.95\pm0.24$ & 142 & $2.14\pm0.11$ & $+1.35\pm0.20$ & $ 48\pm 4$ & $111\pm10$ & $-0.43,-0.76$ & SDSS J095906.47$+$000853.41 \\
   & $218.3\pm10.2$ & 283.9 & 19.829 & $10583\pm1118$ & $4.53\pm0.43$ & 200 & $2.31\pm0.24$ & $+0.93\pm0.41$ & $ 65\pm12$ & $160\pm31$ & $-0.31,-0.81$ & SDSS J101359.79$+$563111.65 \\
   & $504.0\pm 5.1$ & 348.4 & 18.479 & $11230\pm1066$ & $4.24\pm0.44$ & 167 & $2.82\pm0.20$ & $+0.33\pm0.43$ & $ 45\pm 8$ & $ 97\pm21$ & $-0.62,-0.75$ & SDSS J103357.26$-$011507.35 \\
   & $447.7\pm 7.9$ & 294.1 & 19.235 & $10570\pm1443$ & $3.95\pm0.56$ & 302 & $2.95\pm0.26$ & $-0.06\pm0.47$ & $ 74\pm16$ & $178\pm40$ & $-0.31,-0.48$ & SDSS J104318.29$-$013502.51 \\
   & $482.0\pm 4.0$ & 319.6 & 17.381 & $11497\pm 863$ & $3.97\pm0.33$ & 296 & $3.26\pm0.19$ & $-0.32\pm0.41$ & $ 36\pm 6$ & $ 81\pm16$ & $-1.02,-0.76$ & SDSS J112255.77$-$094734.92 \\
   & $414.8\pm 4.6$ & 288.9 & 18.133 & $11087\pm 914$ & $4.55\pm0.42$ & 144 & $2.47\pm0.19$ & $+0.76\pm0.38$ & $ 30\pm 5$ & $ 74\pm13$ & $-1.34,-1.01$ & SDSS J115245.91$-$021116.21 \\
   & $212.8\pm 2.4$ & 289.1 & 17.492 & $11767\pm 744$ & $3.85\pm0.29$ &  43 & $3.56\pm0.19$ & $-0.69\pm0.40$ & $ 43\pm 8$ & $104\pm19$ & $-0.98,-0.64$ & SDSS J140432.38$+$352258.41 \\
   & $280.8\pm 5.7$ & 286.8 & 18.399 & $11170\pm 830$ & $4.92\pm0.32$ &  50 & $2.14\pm0.16$ & $+1.29\pm0.30$ & $ 24\pm 4$ & $ 58\pm 9$ & $-1.82,-1.03$ & SDSS J141723.34$+$101245.74 \\
   & $215.1\pm 7.9$ & 283.9 & 18.884 & $11563\pm1224$ & $4.04\pm0.48$ & 152 & $3.21\pm0.24$ & $-0.23\pm0.46$ & $ 61\pm14$ & $151\pm37$ & $-0.56,-0.48$ & SDSS J154806.92$+$093423.93 \\
   & $ 71.1\pm 9.7$ & 288.3 & 17.510 & $11150\pm1224$ & $3.87\pm0.52$ & 325 & $3.28\pm0.26$ & $-0.44\pm0.47$ & $ 38\pm 8$ & $ 92\pm20$ & $-0.57,+0.16$ & SDSS J180050.86$+$482424.63 \\
   & $130.0\pm10.4$ & 277.8 & 17.354 & $11993\pm1303$ & $4.16\pm0.54$ &  42 & $3.21\pm0.26$ & $-0.13\pm0.48$ & $ 32\pm 7$ & $ 80\pm19$ & $+1.09,-0.88$ & SDSS J232229.47$+$043651.45 \\
        \enddata
\end{deluxetable*}

\clearpage

\end{document}